\documentclass{ws-procs961x669}
\usepackage{subfigure}     
\usepackage{bm}
\usepackage{hyperref}
\usepackage[utf8]{inputenc}

\newcommand{\bra}[1]{\langle #1 |}
\newcommand{\ket}[1]{| #1 \rangle}
\newcommand{\braket}[2]{\langle #1 | #2 \rangle}
\newcommand{\vev}[1]{\langle #1 \rangle}

\begin{document}

\title{Halliday-Suranyi Approach to the Anharmonic Oscillator}

\author{Nabin Bhatta\footnote{nabinb@vt.edu} and Tatsu Takeuchi\footnote{takeuchi@vt.edu}}

\address{Center for Neutrino Physics, Department of Physics\\
Virginia Tech, Blacksburg VA 24061, USA}

\begin{abstract}
In this contribution to \textit{Peter Suranyi Festschrift},
we study the Halliday-Suranyi perturbation method for
calculating the energy eigenvalues of the quartic anharmonic oscillator.
\end{abstract}


\bodymatter

\section{Introduction}

The LHC's non-discovery of new particles that were predicted by proposed solutions to the hierarchy problem
suggests that our understanding of perturbative quantum field theory (QFT) is still limited.
To better understand the behavior of QFT under perturbation theory, it is prudent to go back to the basics
and study the simplest possible case, which
would be interacting bosonic field theory in $0+1$ dimensions, namely,
quantum mechanics (QM) with Hamiltonian
\begin{equation}
\hat{H} \,=\, \dfrac{1}{2}\hat{p}^2 + \dfrac{1}{2}m^2\hat{q}^2 + \dfrac{1}{4}M^3\hat{q}^4\;.
\label{Hamiltonian}
\end{equation}
Here, the operators $\hat{q}$ and $\hat{p}$ have mass dimensions $-\frac{1}{2}$ and $+\frac{1}{2}$, respectively,
and $[\hat{q},\hat{p}]=i$, while $m$ and $M$ both have mass dimension 1.
This Hamiltonian has been studied by many authors since the dawn of QM,
both for practical applications and also as a testbed for various approximation techniques. \cite{Milne_1930,Bell_1945,McWeeny_1948}.

Let us denote the eigenvalues of $\hat{H}$ by $E_n(m,M)$, $n=0,1,2,\cdots$.
We all know that when $M=0$ we have
\begin{equation}
E_n(m,0) \;=\; m\left(n+\frac{1}{2}\right)\;.
\end{equation}
Treating the harmonic oscillator part of $\hat{H}$ as the unperturbed
Hamiltonian and the quartic part of the potential as the perturbation, i.e.
\begin{equation}
\hat{H}_0 \,=\, \dfrac{1}{2}\hat{p}^2 + \dfrac{1}{2}m^2\hat{q}^2\;,\qquad
\hat{V} \,=\, \dfrac{1}{4}M^3\hat{q}^4\;,
\end{equation}
Rayleigh-Schr\"odinger perturbation theory\cite{Schrodinger:1926vbi} gives
the value of $E_n(m,M)$ as a power series in $\lambda = (M/m)^3$ :
\begin{equation}
E_n(m,M) \;=\; m\bigg[ c_0(n)
+ \lambda\, c_1(n) + \lambda^2 c_2(n) + \lambda^3 c_3(n) + \cdots
\bigg]
\;,
\end{equation}
where
\begin{eqnarray}
c_0(n) & = & n+\frac{1}{2}\;,\vphantom{\bigg|}\cr
c_1(n) & = & \dfrac{3(2n^2+2n+1)}{16}\;,\vphantom{\bigg|}\cr
c_2(n) & = & -\dfrac{34n^3+51n^2+59n+21}{128}\;,\vphantom{\bigg|}\cr
c_3(n) & = & \dfrac{3(125n^4+250n^3+472n^2+347n+111)}{1024}\;,\vphantom{\bigg|}\cr
& \vdots & 
\end{eqnarray}
However, this is a divergent asymptotic series \cite{Kato:1949,Kato:1950a,Kato:1950b,Dyson:1952tj,Jaffe:1965,Bender:1969si,Bender:1971gu,Bender:1973rz,Lipatov:1976ny,Parisi:1976zh} and
must be Borel summed to recover the values of $E_n(m,M)$. \cite{Loeffel:1969rdm,Simon:1970xb,Graffi:1970erh}

On the other hand, when $m=0$ we can argue on dimensional grounds that
\begin{equation}
E_n(0,M) \;=\; M\,A_0(n)\;,
\end{equation}
where $A_0(n)$ is a dimensionless function of $n$, which
scales as $\sim n^{4/3}$ as $n\to\infty$.\cite{Milne_1930,Bell_1945,Hioe:1978jj}
If we treat the quadratic part of the potential as the perturbation instead of the quartic part,
that is: 
\begin{equation}
\hat{H}_0 \,=\, \dfrac{1}{2}\hat{p}^2 + \dfrac{1}{4}M^3\hat{q}^4\;,\qquad
\hat{V} \,=\, \dfrac{1}{2}m^2\hat{q}^2\;,
\end{equation}
then the $m\neq 0$ case is 
expandable in powers of $\lambda^{-2/3}=(m/M)^2$ :
\begin{equation}
E_n(m,M) \;=\; M\bigg[ 
A_0(n) + \dfrac{A_1(n)}{\lambda^{2/3}} + \dfrac{A_2(n)}{\lambda^{4/3}} + \cdots
\bigg]
\;.
\label{SCexpansion}
\end{equation}
This strong-coupling expansion is convergent for $\lambda\gg 1$.\cite{Parisi:1976zh} 
However, the perturbative calculation
of the coefficients $A_k(n)$ is difficult due to the the unperturbed Hamiltonian $\hat{H}_0$
lacking in simple analytic expressions for its eigenvalues and eigenfunctions.\cite{Liverts:2006qi}
Bender et al. in Ref.~\citenum{Bender:1978ew} approach this problem by treating the quartic potential part
as the unperturbed Hamiltonian and the harmonic oscillator part the perturbation, i.e.
\begin{equation}
\hat{H}_0 \;=\; \frac{1}{4}M^3\hat{q}^4\;,\qquad
\hat{V} \;=\; \frac{1}{2}\hat{p}^2 + \dfrac{1}{2}m^2\hat{q}^2\;.
\end{equation}
However, this method requires the introduction of a spatial lattice to regulate the
$\hat{q}^4$ operator, and this lattice spacing must be extrapolated to zero at the end of the calculation.

\section{The Halliday-Suranyi Approach}

In Refs.~\citenum{Halliday:1979vn} and \citenum{Halliday:1979xh}, Halliday and Suranyi introduce
an interesting method for dealing with the quartic anharmonic oscillator.
First, note that the $\hat{q}^4$ operator can be rewritten as
\begin{eqnarray}
\dfrac{1}{4}\,\hat{q}^4
& = & \dfrac{1}{\Omega^4}
\left(\dfrac{1}{2}\Omega^2\hat{q}^2\right)^2
\vphantom{\Bigg|}\cr
& = & \dfrac{1}{\Omega^4}
\left\{
\left( \dfrac{\hat{p}^2}{2} + \dfrac{1}{2}\Omega^2\hat{q}^2 \right)^2
-\dfrac{\Omega^2}{4}
\left(\hat{p}^2\hat{q}^2 + \hat{q}^2\hat{p}^2
\right)
- \dfrac{\hat{p}^4}{4}
\right\}
\;,
\vphantom{\Bigg|}
\end{eqnarray}
where $\Omega$ is an arbitrary mass parameter.
This allows us to separate $\hat{H}$ into the unperturbed Hamiltonian $\hat{H}_0$
and the perturbation $\hat{V}$ as follows:
\begin{eqnarray}
\hat{H} 
& = & \dfrac{1}{4}M^3\hat{q}^4
+ \left(\dfrac{\hat{p}^2}{2} + \dfrac{1}{2}m^2\hat{q}^2\right)
\vphantom{\Bigg|}\cr
& = & 
\underbrace{\dfrac{M^3}{\Omega^4}\left( \dfrac{\hat{p}^2}{2} + \dfrac{1}{2}\Omega^2\hat{q}^2 \right)^2}_{\displaystyle \hat{H}_0}
+\underbrace{
\left(\dfrac{\hat{p}^2}{2} + \dfrac{1}{2}m^2\hat{q}^2\right)
-\dfrac{M^3}{\Omega^4}
\left\{\dfrac{\Omega^2}{4}
\left(\hat{p}^2\hat{q}^2 + \hat{q}^2\hat{p}^2
\right)
+ \dfrac{\hat{p}^4}{4}
\right\}
}_{\displaystyle \hat{V}}
\;.\cr
& & 
\end{eqnarray}
Note that by the replacement
\begin{equation}
\dfrac{1}{4}M^3\hat{q}^4 \quad\to\quad
\dfrac{M^3}{\Omega^4}\left( \dfrac{\hat{p}^2}{2} + \dfrac{1}{2}\Omega^2\hat{q}^2 \right)^2
\end{equation}
we discretize the eigenvalues $\hat{H}_0$ with $\Omega$ acting as the regulator,
without the introduction of a spatial lattice.
The eigenvalues of $\hat{H}_0$ in units of $M$ are
\begin{equation}
E_n^{(0)}(Z) 
\;=\; \dfrac{M}{Z^{2/3}}\left(n +\dfrac{1}{2}\right)^2
\;,\qquad
n\;=\;0,1,2,\cdots\;,
\end{equation}
where $Z=(\Omega/M)^3$.
Denote the expansion of $E_n(m,M)$ in powers of $\hat{V}$ as
\begin{eqnarray}
E_n(m,M) & = & E_n^{(0)}(Z) + E_n^{(1)}(Z) + E_n^{(2)}(Z) + E_n^{(3)}(Z) + \cdots
\label{HSexpansion}
\end{eqnarray}
The convergence of this series is demonstrated in Ref.~\citenum{Halliday:1979xh}.
The first few terms of this expansion are given by
\begin{eqnarray}
\lefteqn{E_n^{(1)}(Z) 
\;=\; \dfrac{M}{Z^{2/3}}\bigg(
\dfrac{2n+1}{4}Z(1\!+\!X)
-\dfrac{10n^2 + 10n + 1}{16}
\bigg)
\;,
}
\vphantom{\Bigg|}\cr
\lefteqn{E_n^{(2)}(Z) 
\;=\; -\dfrac{M}{Z^{2/3}}
\Bigg[
\dfrac{3(n^4+2n^3-2n^2-3n-3)}{2^7(2n-3)(2n+5)}
}
\vphantom{\Bigg|}\cr
& & 
-\dfrac{n(n-1)}{2^5(2n-1)}\bigg(Z(1\!-\!X) - \dfrac{2n-1}{2}\bigg)^2 
+\dfrac{(n+1)(n+2)}{2^5(2n+3)}\bigg(Z(1\!-\!X) - \dfrac{2n+3}{2}\bigg)^2
\Bigg]
\;,
\vphantom{\Bigg|}\cr
\lefteqn{E_n^{(3)}(Z) 
\;=\; \dfrac{M}{Z^{2/3}}\Bigg[
}\vphantom{\Bigg|}\cr
& & 
\dfrac{(n+1)(n+2)}{4(2n+3)^2}
\bigg(\dfrac{Z(1-X)}{4} - \dfrac{2n+3}{8}\bigg)^2
\bigg(\dfrac{2n+5}{4}Z(1\!+\!X) - \dfrac{10n^2+50n+61}{16}\bigg)
\vphantom{\Bigg|}\cr
& & 
+\dfrac{n(n-1)}{4(2n-1)^2}
\bigg(\dfrac{Z(1-X)}{4}-\dfrac{2n-1}{8}\bigg)^2
\bigg(\dfrac{2n-3}{4}Z(1\!+\!X) - \dfrac{10n^2-30n+21}{16}\bigg)
\vphantom{\Bigg|}\cr
& & 
+\dfrac{(n+1)(n+2)(n+3)(n+4)}{2^{6}(2n+5)(2n+3)}
\bigg(\dfrac{Z(1-X)}{4}-\dfrac{2n+3}{8}\bigg)
\bigg(\dfrac{Z(1-X)}{4}-\dfrac{2n+7}{8}\bigg)
\vphantom{\Bigg|}\cr
& & 
-\dfrac{n(n-1)(n+1)(n+2)}{2^{5}(2n+3)(2n-1)}
\bigg(\dfrac{Z(1-X)}{4}-\dfrac{2n-1}{8}\bigg)
\bigg(\dfrac{Z(1-X)}{4}-\dfrac{2n+3}{8}\bigg)
\vphantom{\Bigg|}\cr
& & 
+\dfrac{n(n-1)(n-2)(n-3)}{2^{6}(2n-1)(2n-3)}
\bigg(\dfrac{Z(1-X)}{4}-\dfrac{2n-5}{8}\bigg)
\bigg(\dfrac{Z(1-X)}{4}-\dfrac{2n-1}{8}\bigg)
\vphantom{\Bigg|}\cr
& &
+\dfrac{(n+1)(n+2)(n+3)(n+4)}{2^{12}(2n+5)^2}
\bigg(\dfrac{2n+9}{4}Z(1+X) - \dfrac{10n^2+90n+201}{16}\bigg)
\vphantom{\Bigg|}\cr
& & 
+\dfrac{n(n-1)(n-2)(n-3)}{2^{12}(2n-3)^2}
\bigg(\dfrac{2n-7}{4}Z(1+X)- \dfrac{10n^2-70n+121}{16}\bigg)
\vphantom{\Bigg|}\cr
& & 
-\Bigg\{
\dfrac{(n+1)(n+2)}{4(2n+3)^2}
\bigg(\dfrac{Z(1-X)}{4}-\dfrac{2n+3}{8}\bigg)^2
+\dfrac{(n+1)(n+2)(n+3)(n+4)}{2^{12}(2n+5)^2}
\vphantom{\Bigg|}\cr
& & \quad
+\dfrac{n(n-1)}{4(2n-1)^2}
\bigg(\dfrac{Z(1-X)}{4}-\dfrac{2n-1}{8}\bigg)^2
+\dfrac{n(n-1)(n-2)(n-3)}{2^{12}(2n-3)^2}
\Bigg\}
\vphantom{\Bigg|}\cr
& & \quad\times
\bigg(\dfrac{2n+1}{4}Z(1+X)-\dfrac{10n^2+10n+1}{16}\bigg)
\Bigg]
\;,\vphantom{\Bigg|}
\label{HSterms}
\end{eqnarray}
where we have used the shorthand
\begin{equation}
X\;=\;\dfrac{m^2}{\Omega^2} \;=\; \dfrac{1}{(\lambda Z)^{2/3}}\;.
\end{equation}
Collecting the powers of $X$ would lead to the strong coupling expansion of \eref{SCexpansion}.
If we set $n=0$ and $X=0$ (i.e. $m=0$) in the above expressions, we recover
Eq.~(2.6) of Ref.~\citenum{Halliday:1979xh}.

\section{Choice of the parameter $Z$}

\begin{figure}[t]
\centerline{
\includegraphics[width=12cm]{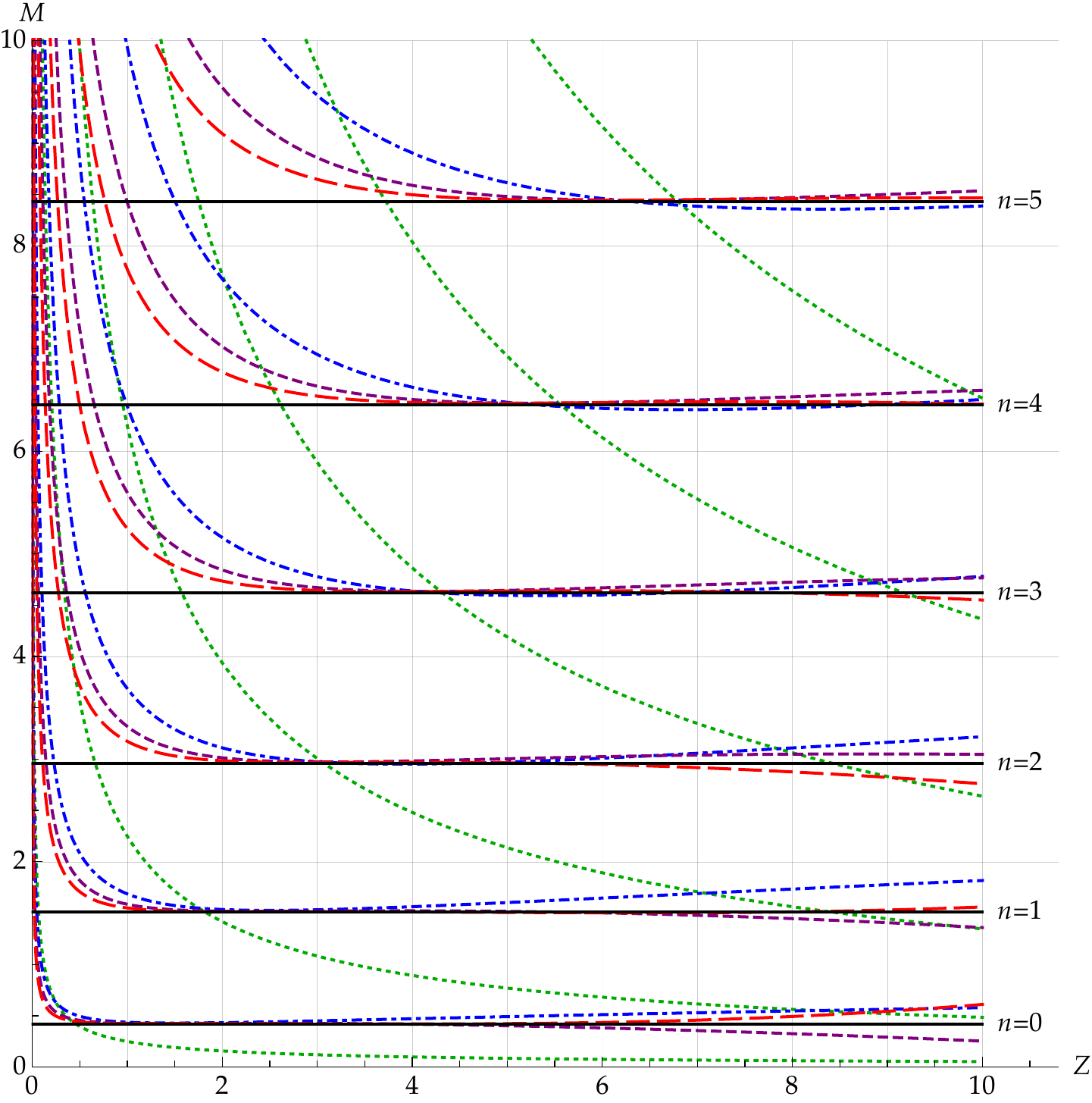}
}
\caption{The Halliday-Suranyi expansion for the $m=0$ case compared with the exact result
for the states $n=0$ through $n=5$.
Dotted line: 0th order, dot-dashed line: 1st order, short-dashed line: 2nd order,
long-dashed line: 3rd order, solid horizontal line: exact value.
We can see that the optimum value of $Z$ for state $n$ is $Z\sim n+1$.
}
\label{HSfig}
\end{figure}

Note that though every term in the expansion of \eref{HSexpansion} depends 
on $Z = (\Omega/M)^3$, the sum that the series converges to does not since the full Hamiltonian
$\hat{H}$ is independent of the arbitrary parameter $\Omega$ used to 
separate $\hat{H}$ into $\hat{H}_0$ and $\hat{V}$.
However, when the series is truncated after a finite number of terms, the dependence 
on $Z$ will remain.
This is illustrated for the $m=0$ case in \fref{HSfig}, in which the
exact numerical results for $A_0(n)$, $n=0,1,\cdots,5$, are compared to the 0th, 1st, 2nd, and 3rd order approximations.
From \fref{HSfig}, it is evident that for each $n$ there is an optimum value of $Z$ for which the series
converges quickly and the first few terms provide a very good approximation.
By inspection, we expect this value to scale as
\begin{equation}
Z \;\sim\; n+1\;.
\end{equation}
The question is: what is the best procedure to fix $Z$ so that the resulting approximation is good?
Note that the problem is similar to the renormalization scale setting problem in 
perturbative QCD and, consequently, we borrow some of the language used in that field.\cite{Wu:2013ei}


\newpage
\subsection{Method 1 : Fastest Apparent Convergence}

\begin{figure}[b]
\centerline{
\subfigure[]{\includegraphics[height=4cm]{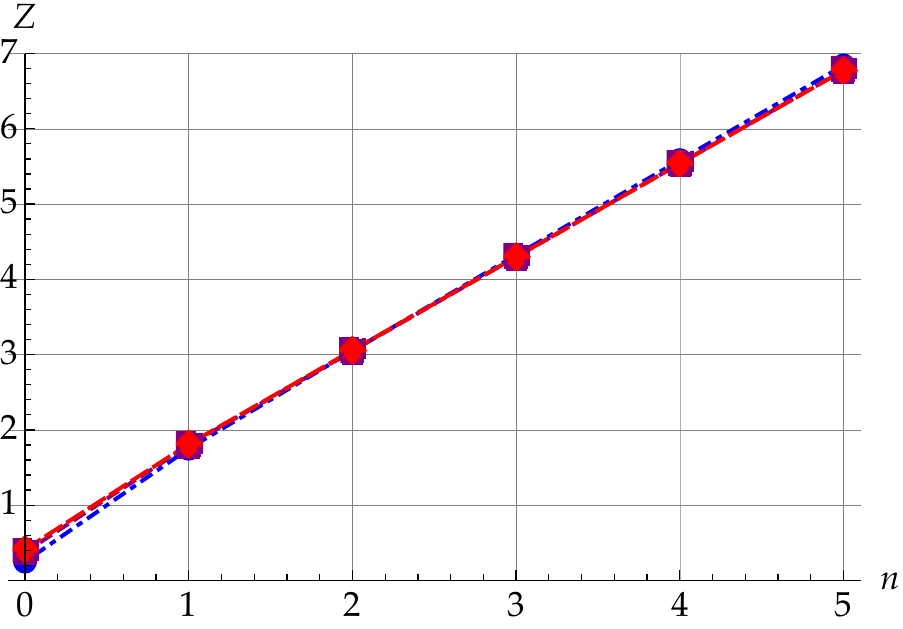}\label{AP1}}
\subfigure[]{\includegraphics[height=4cm]{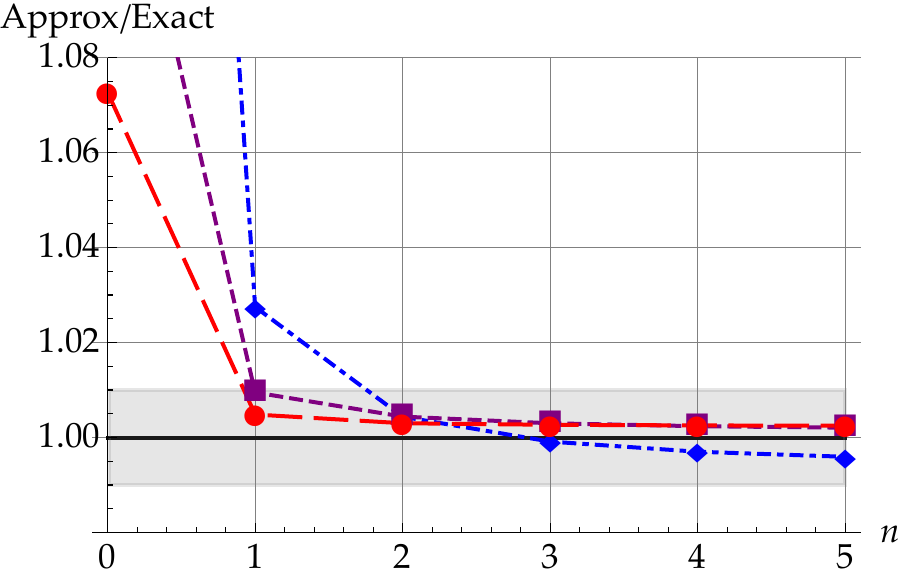}\label{AP2}}
}
\caption{(a) Mimimim values of $Z$ that solve \eref{ZPcondition} for the $X=0$ case, and
(b) approximate over exact values of $E_n(0,M)$ at those $Z$.
Results are shown for $k=1$ (diamonds, dotdashed), $k=2$ (squares, short-dashed), and $k=3$ (circles, long-dashed).
The three lines are overlapping in (a) and difficult to distinguish.}
\label{AP}
\end{figure}

In Ref.~\citenum{Halliday:1979xh},
Halliday and Suranyi consider demanding that
\begin{equation}
E_n(m,M) \;\approx\; 
E_n^{(0)}(Z)
+ \underbrace{E_n^{(1)}(Z) + E_n^{(2)}(Z) + \cdots + E_n^{(k)}(Z)}_{\displaystyle =0}
\label{FAC}
\end{equation}
to fix $Z$ at each order $k$. 
This corresponds to the method of Fastest Apparent Convergence (FAC) used in pQCD.\cite{Wu:2013ei}
For $k=1$, we need to solve
\begin{equation}
E_n^{(1)}(Z) \,=\, 0 
\quad\to\quad
Z + \frac{m^2}{M^2}Z^{1/3} - \dfrac{5\left(n^2+n+\frac{1}{10}\right)}{4\left(n+\frac{1}{2}\right)} \,=\, 0 \;,
\end{equation}
which, in general, has one real and two complex solutions.
For the $m=0$ (i.e. $X=0$) case, the three solutions overlap and we have
\begin{equation}
Z \;=\;  Z_{n,\text{FAC}}^{(1)} 
\;\equiv\; \dfrac{5\left(n^2+n+\frac{1}{10}\right)}{4\left(n+\frac{1}{2}\right)}
\;\xrightarrow{n\gg 1}\; \frac{5}{4}\bigg(n+\frac{1}{2}\bigg)
\;.
\label{ZPapprox}
\end{equation}
The value of $E_n^{(0)}(Z)$ at $Z=Z_{n,\text{FAC}}^{(1)}$ is
\begin{eqnarray}
E_n^{(0)}(Z_{n,\text{FAC}}^{(1)}) & = &
M\bigg(\dfrac{4}{5}\bigg)^{2/3}
\bigg[
\dfrac{(n+\frac{1}{2})^4}{(n^2+n+\frac{1}{10})}
\bigg]^{2/3}
\vphantom{\Bigg|}\cr
& \xrightarrow{n\gg 1} &
M\bigg(\dfrac{4}{5}\bigg)^{2/3}
\bigg(n+\frac{1}{2}\bigg)^{4/3}
=\; 0.862\,M\bigg(n+\frac{1}{2}\bigg)^{4/3}
\;.
\end{eqnarray}
Note that this expression scales as $\sim n^{4/3}$ for large $n$ as it should.

Graphically, \eref{FAC} is equivalent to searching for values of $Z$ at which
the graphs for the 0th, and $k$th order approximations cross:
\begin{equation}
E_n^{(0)}(Z) \;=\; \sum_{j=0}^{k}E_n^{(j)}(Z) \;.
\label{ZPcondition}
\end{equation}
%

\begin{figure}[h!]
\centerline{
\includegraphics[height=16cm]{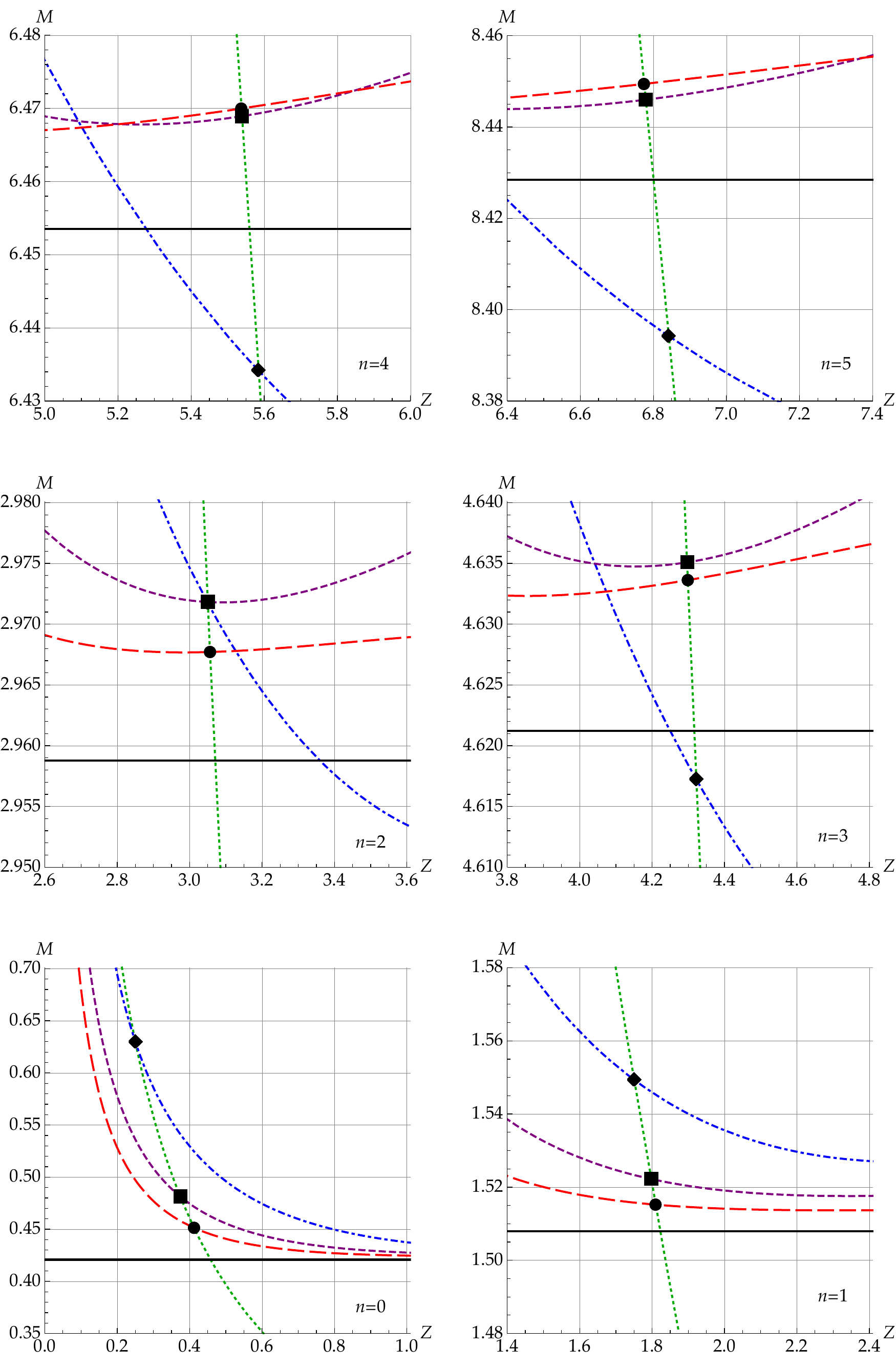}
}
\caption{Blowups of \fref{HSfig} showing the crossing points of
the 0th order (dotted) with the 1st (dot-dashed), 2nd (short-dashed),
and 3rd (long-dashed) order approximations.
The 2nd and 3rd order lines can have multiple crossings with the 0th order line.
Here, we show the crossings with the smallest values of $Z$. 
}
\label{FACfig}
\end{figure}

\noindent
This is illustrated for the $X=0$ case in
\fref{FACfig}.
The six graphs shown are for the $n=0$ to $n=5$ states,
and each shows the intersections of the $k=0$ graph (dotted) with the
$k=1$ (dotdashed), $k=2$ (short-dashed), and $k=3$ (long-dashed) graphs.
One problem with this approach is that there exist, in general,
multiple real solutions to \eref{ZPcondition} for $k\ge 2$.
From these multiple real solutions, we choose the smallest $Z$ for each $k$.
This gives us the crossing point closest to the vertical axis, which are the ones
shown in \fref{FACfig}. 
The values of $Z$ and $E_n^{(0)}(Z)$ at these points are graphed in
\fref{AP1} and \fref{AP2}, respectively.
We can see from \fref{AP1} that the $Z$ values for $k\ge 2$ never deviate away from the $k=1$ case \eref{ZPapprox},
and from \fref{AP2} that, for $n\ge 2$, the $k=1$ value is already within 1\% of the actual value.

\subsection{Method 2 : Principle of Minimum Sensitivity}

Another method for choosing $Z$ would be to require
\begin{equation}
\dfrac{d}{dZ}\bigg[\sum_{j=0}^{k} E_n^{(j)}(Z) \bigg] \;=\; 0\;.
\label{PMScondition}
\end{equation}
This corresponds to the Principle of Minimum Sensitivity (PMS) used in pQCD.\cite{Wu:2013ei}
For the $k=1$ case, PMS is equivalent to minimizing the expectation value of $\hat{H} = \hat{H}_0 + \hat{V}$
using the harmonic oscillator eigenfunctions as the variational trial functions.\cite{Weinstein:2005kx}
Indeed, $E_n^{(0)}(Z)$ and $E_n^{(1)}(Z)$ are respectively the expectation values of
$\hat{H}_0$ and $\hat{V}$ for the $n$th eigenstate of $\hat{H}_0$.
Imposing \eref{PMScondition} for $k=1$, we find
\begin{eqnarray}
\dfrac{d}{dZ}\Big[
E_n^{(0)}(Z) + E_n^{(1)}(Z)
\Big]
& = & 0 \cr
& \downarrow & \cr
Z - \frac{m^2}{M^2}Z^{1/3} - \dfrac{3\left(n^2+n+\frac{1}{2}\right)}{2\left(n+\frac{1}{2}\right)} 
& = & 0 
\;,
\end{eqnarray}
which, in general, has one real and two complex solutions. 
For the $m=0$ (i.e. $X=0$) case, we find
\begin{equation}
Z \;=\; Z_{n,\text{PMS}}^{(1)} 
\;\equiv\; \dfrac{3\left(n^2+n+\frac{1}{2}\right)}{2\left(n+\frac{1}{2}\right)} 
\;\xrightarrow{n\gg 1}\;\dfrac{3}{2}\bigg(n+\frac{1}{2}\bigg)\;,
\end{equation}
and the approximate value at $Z=Z_{n,\text{PMS}}^{(1)}$ is \cite{Weinstein:2005kx}
\begin{eqnarray}
\lefteqn{E_n^{(0)}(Z_{n,\text{PMS}}^{(1)}) + E_n^{(1)}(Z_{n,\text{PMS}}^{(1)})}
\vphantom{\bigg|}\cr
& = & M\,
\dfrac{3^{4/3}}{2^{7/3}}
\bigg[
\bigg(n+\frac{1}{2}\bigg)^2
\bigg(n^2+n+\frac{1}{2}\bigg)
\bigg]^{1/3}
\vphantom{\Bigg|}\cr
& \xrightarrow{n\gg 1} &
M\,\dfrac{3^{4/3}}{2^{7/3}}
\bigg(n+\frac{1}{2}\bigg)^{4/3}
\;=\; 0.859\,M
\bigg(n+\frac{1}{2}\bigg)^{4/3}
\;,
\label{RGapprox}
\end{eqnarray}
which is numerically similar to \eref{ZPapprox}.
For $k\ge 2$, PMS does not correspond to any variational calculation.

\begin{figure}[h!]
\centerline{
\includegraphics[height=16cm]{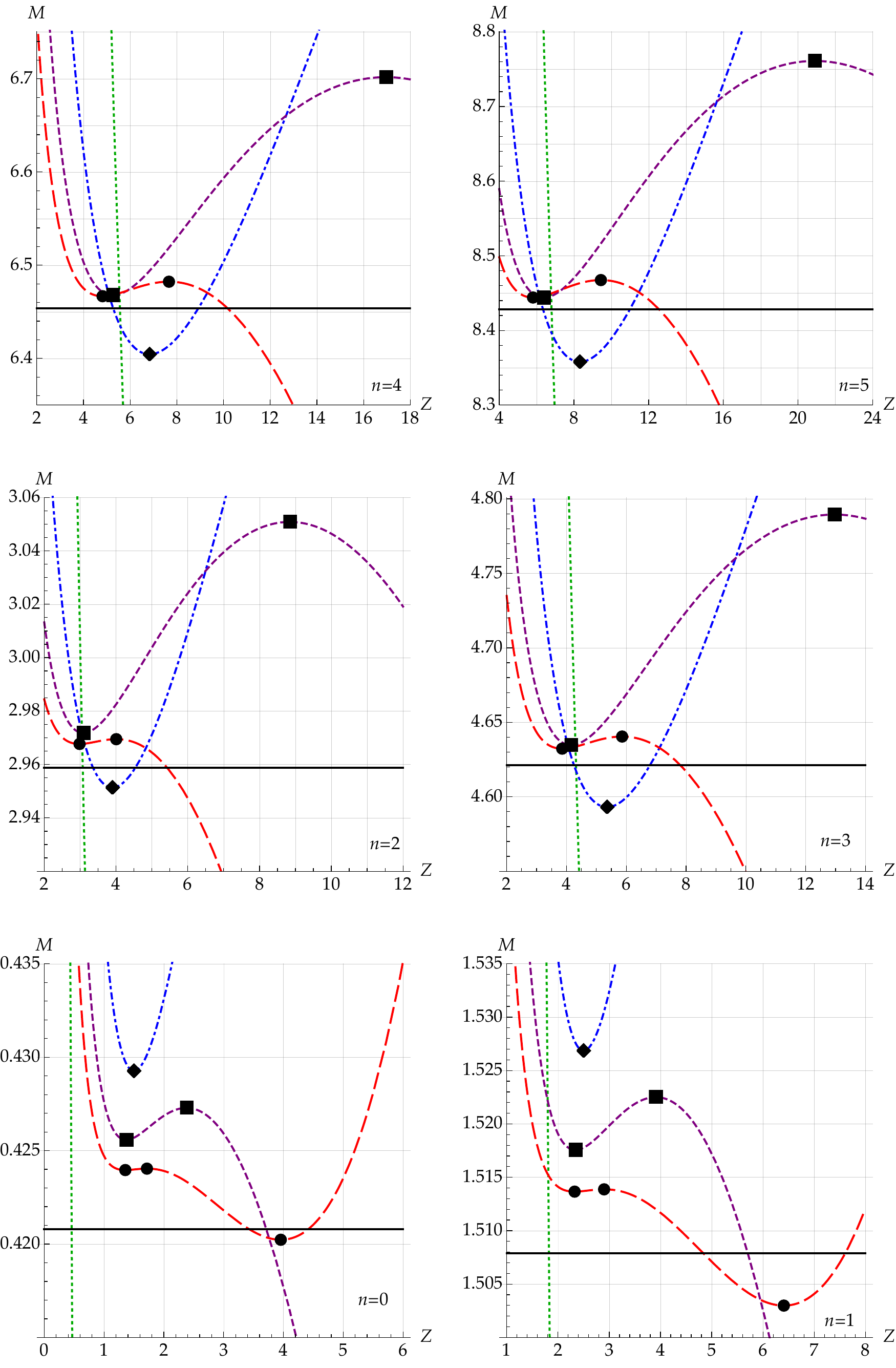}
}
\caption{The Principle of Minimum Sensitivity (PMS) applied to the states $n=0$ through $n=5$.
The diamonds, squared, and circles respectively show the points at which the
1st (dot-dashed), 2nd (short-dashed), and 3rd (long-dashed) order approximations
are flat.
The 2nd order graph has two flat points, with the left point giving a local minimum,
and the right point a local maximum. 
The 3rd order graph has three flat points, with the left-most and right-most points giving local
minima, and the middle point a local maximum.
However, the right local minimum always undershoots the exact value for all $n$, and dips into the negative for $n\ge 2$.
}
\label{PMSfig}
\end{figure}

Graphically, \eref{PMScondition} looks for the values of $Z$ for which the slope
of the $k$th order approximation is flat.
This is illustrated for the $X=0$ case in \fref{PMSfig}, in which
the six graphs shown are for the $n=0$ to $n=5$ states.
For $k=1$, there is a unique flat location as we saw above.
For $k=2$, there are two flat locations in which the one on the left is a 
local minimum while the one of the right is a local maximum.
Though we cannot tell which one should be choosen beforehand,
comparison with the exact results suggests we should choose the local minimum 
point on the left. 

For $k=3$, there are three flat locations, where the two outer points
are local minima, while the one in the middle is a local maximum.
For all the cases considered, the left local minimum is closer to the exact result
than the central local maximum.
For the ground state, $n=0$, the right local minimum is closer to the exact result
than the left local minimum, but undershoots it.
For $n\ge 2$, though we cannot tell from \fref{PMSfig}, the
right local minimum dips into the negative.
Thus, we choose the left local minimum as our approximation for $E_n(0,M)$.

\begin{figure}[t]
\centerline{
\subfigure[]{\includegraphics[height=4cm]{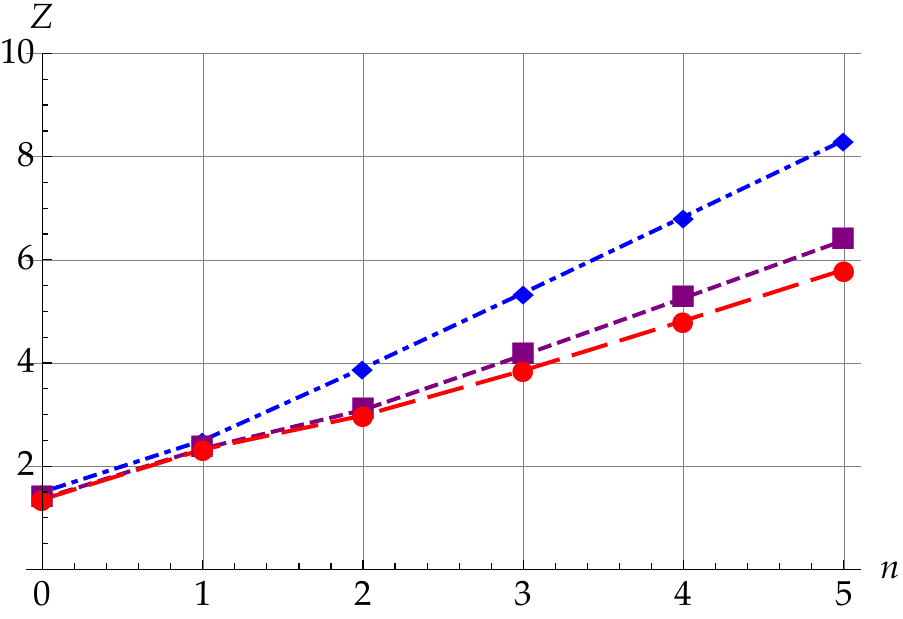}\label{PMS1}}
\subfigure[]{\includegraphics[height=4cm]{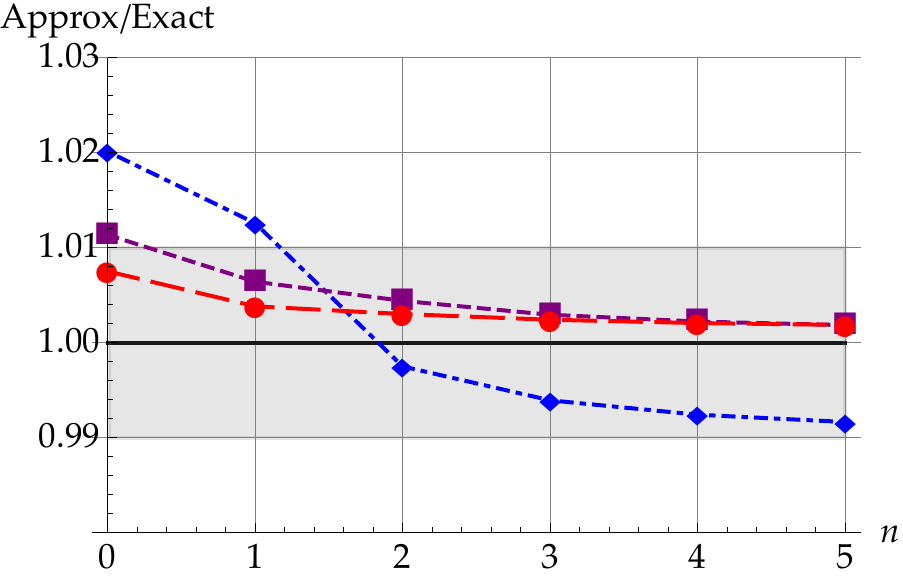}\label{PMS2}}
}
\caption{(a) Mimimim values of $Z$ that solve \eref{PMScondition} for the $X=0$ case, and
(b) approximate over exact values of $E_n(0,M)$ at those $Z$.
Results are shown for $k=1$ (diamonds, dotdashed), $k=2$ (squares, short-dashed), and $k=3$ (circles, long-dashed).
}
\label{PMSza}
\end{figure}

The values of $Z$ chosen in this way are plotted in \fref{PMS1}, and the
resulting approximate values normalized to the exact value are shown in \fref{PMS2}.
At $k=3$, the approximate values are within 1\% of the exact value for all $n$ 
considered.

\subsection{Method 3 : Perturbative Variational Method}

The problem with \eref{ZPcondition} and \eref{PMScondition} is that they do not uniquely determine
$Z$ for $k\ge 2$, and choosing the smallest real solution was somewhat arbitrary and there is no 
guarantee that this choice would be optimal.
To remedy this problem, let us consider the following.
Denote the perturbative expansion of the eigenstates of $\hat{H}$ in powers of $\hat{V}$ as
\begin{equation}
\ket{n} \;=\; \ket{n^{(0)}} + \ket{n^{(1)}} + \ket{n^{(2)}} + \ket{n^{(3)}} + \cdots\;,
\end{equation}
where $\ket{n^{(k)}}$ includes all terms proportional to $k$ powers of $\hat{V}$.
We have commented in the previous subsection that
\begin{eqnarray}
\vev{H}_n^{(0)}
\; \equiv \; \dfrac{\bra{n^{(0)}}\hat{H}\ket{n^{(0)}}}{\braket{n^{(0)}}{n^{(0)}}}
& = & \bra{n^{(0)}}\Big(\hat{H}_0+\hat{V}\Big)\ket{n^{(0)}}
\vphantom{\bigg|}\cr
& = &
 \underbrace{\bra{n^{(0)}}\hat{H}_0\ket{n^{(0)}}}_{\displaystyle E_n^{(0)}}
+\underbrace{\bra{n^{(0)}}\hat{V}\ket{n^{(0)}}}_{\displaystyle E_n^{(1)}}
\;,
\end{eqnarray}
so the PMS condition, \eref{PMScondition}, applied to the $k=1$ case minimizes $\vev{H}_n^{(0)}$ using $\ket{n^{(0)}}$ as
the trial function.
Now, consider the expectation value of $\hat{H}$ for the state $\ket{n^{0}}+\ket{n^{(1)}}$ :
\begin{equation}
\vev{H}_n^{(1)}
\; \equiv \;
\dfrac{\big(\bra{n^{(0)}}+\bra{n^{(1)}}\big)\big(\hat{H}_0+\hat{V}\big)\big(\ket{n^{0}}+\ket{n^{(1)}}\big)}
{\big(\bra{n^{(0)}}+\bra{n^{(1)}}\big)\big(\ket{n^{0}}+\ket{n^{(1)}}\big)}
\;.
\label{H1def}
\end{equation}
Minimizing $\vev{H}_n^{(1)}$ by varying  $\ket{n^{0}}+\ket{n^{(1)}}$ 
should improve the approximations for the $n=0$ (ground state, even parity)
and the $n=1$ (1st excited state, odd parity) cases without ever going below the exact values.

The numerator and denominator of \eref{H1def} are
\begin{eqnarray}
\lefteqn{
\big(\bra{n^{(0)}}+\bra{n^{(1)}}\big)\big(\hat{H}_0+\hat{V}\big)\big(\ket{n^{0}}+\ket{n^{(1)}}\big)
}
\vphantom{\bigg|}\cr
& = &  \underbrace{\bra{n^{(0)}}\hat{H}_0\ket{n^{(0)}}}_{\displaystyle E_n^{(0)}}
+\underbrace{\bra{n^{(0)}}\hat{V}\ket{n^{(0)}}}_{\displaystyle E_n^{(1)}}
+\underbrace{\bra{n^{(0)}}\hat{H}_0\ket{n^{(1)}}+\bra{n^{(1)}}\hat{H}_0\ket{n^{(0)}}}_{\displaystyle 0}
\cr
& & + \underbrace{\bra{n^{(0)}}\hat{V}\ket{n^{(1)}} + \bra{n^{(1)}}\hat{V}\ket{n^{(0)}}}_{\displaystyle
2E_n^{(2)}
}
\cr
& &
+ \underbrace{\bra{n^{(1)}}\hat{H}_0\ket{n^{(1)}}}_{\displaystyle
-E_n^{(2)} + E_n^{(0)}\braket{n^{(1)}}{n^{(1)}}
} + 
\underbrace{\bra{n^{(1)}}\hat{V}\ket{n^{(1)}}}_{\displaystyle
E_n^{(3)}+E_n^{(1)}\braket{n^{(1)}}{n^{(1)}}
}
\cr
& = & \big(E_n^{(0)}+E_n^{(1)}\big)\big(1 + \braket{n^{(1)}}{n^{(1)}}\big)
+ E_n^{(2)} + E_n^{(3)}
\;,
\vphantom{\bigg|}\cr
\lefteqn{
\big(\bra{n^{(0)}}+\bra{n^{(1)}}\big)\big(\ket{n^{0}}+\ket{n^{(1)}}\big)
} 
\vphantom{\bigg|}\cr
& = & \underbrace{\braket{n^{(0)}}{n^{(0)}}}_{\displaystyle 1}
+ \underbrace{\braket{n^{(0)}}{n^{(1)}}+\braket{n^{(1)}}{n^{(0)}}}_{\displaystyle 0}
+ \braket{n^{(1)}}{n^{(1)}}
\vphantom{\bigg|}\cr
& = & 1 + \braket{n^{(1)}}{n^{(1)}}
\;.
\vphantom{\bigg|}
\end{eqnarray}
Therefore,
\begin{equation}
\vev{H}_n^{(1)}
\;=\; E_n^{(0)} + E_n^{(1)} + \dfrac{E_n^{(2)} + E_n^{(3)}}{1 + \braket{n^{(1)}}{n^{(1)}}}
\;.
\end{equation}
We see that $\vev{H}_n^{(1)}$ can be considered the 3rd-order perturbative energy $\sum_{j=0}^{3}E_n^{(j)}$
corrected by a class of higher order terms which have been resummed into the factor
$(1+\braket{n^{(1)}}{n^{(1)}})^{-1}$.
This is similar in spirit to Renormalization Group resummation, or the
Brodsky-Lepage-Mackenzie method used in pQCD.\cite{Wu:2013ei}
The correction renders $\vev{H}_n^{(1)}$ positive definite, and avoids the problem 
$\sum_{j=0}^{3}E_n^{(j)}$ has of becoming negative around the right local minimum for $n\ge 2$.


\begin{figure}[h!]
\centerline{
\includegraphics[height=16cm]{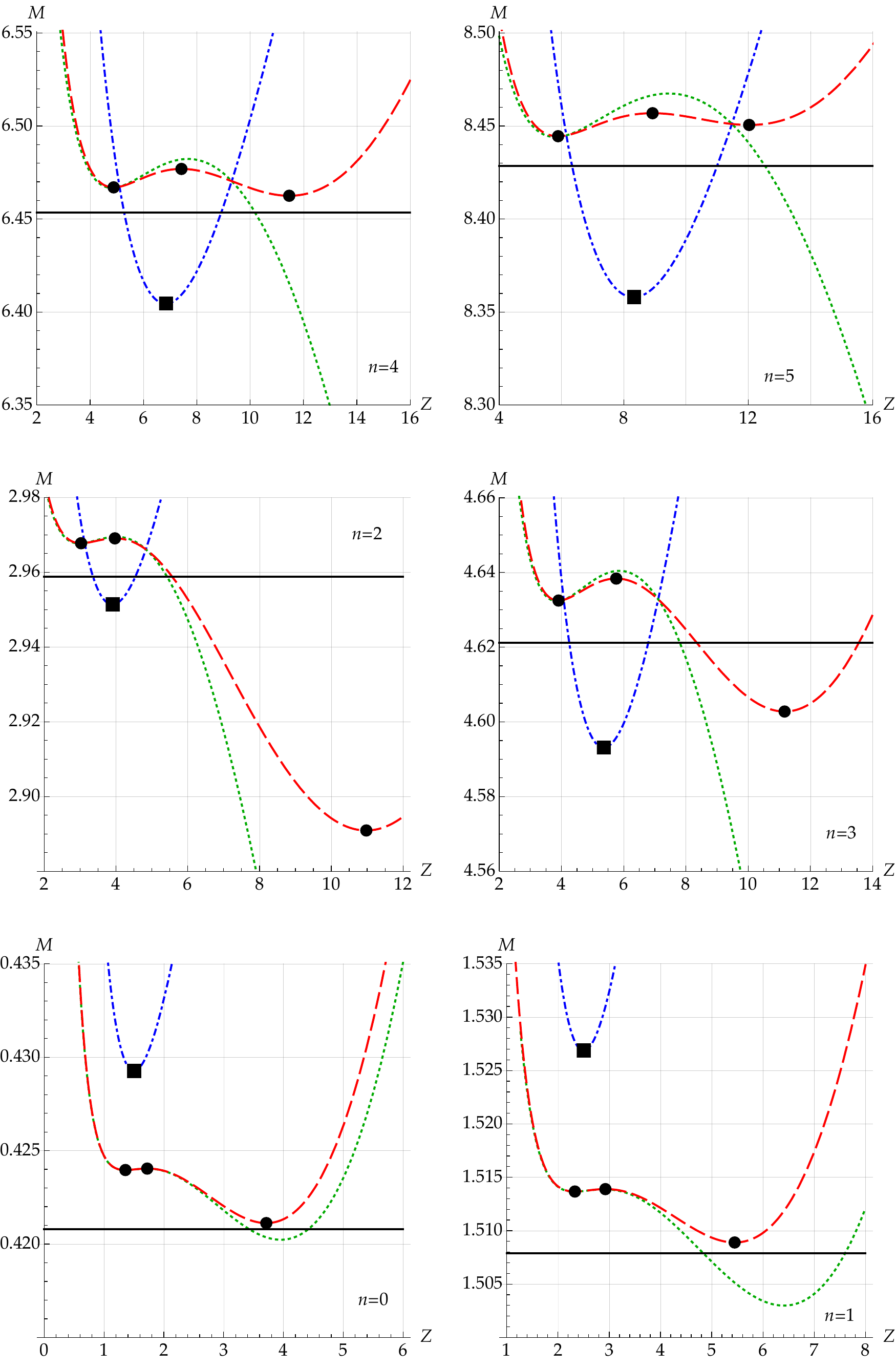}
}
\caption{Stationary points of $\vev{H}_n^{(0)} = E_n^{(0)} + E_n^{(1)}$ (squares)
and $\vev{H}_n^{(1)}$ (circles) for $n=0$ through $n=5$.
The graphs shown are those for $\vev{H}_n^{(0)}$ (dot-dashed), $\vev{H}_n^{(1)}$ (dashed), and
$\sum_{j=0}^{3}E_n^{(j)}$ (dotted).
Solid horizontal lines indicate the exact values.
The stationary points of $\vev{H}_n^{(0)}$ are the same as the $k=1$ PMS points in \fref{PMSfig}.
The right local minimum of $\sum_{j=0}^{3}E_n^{(j)}$ has been lifted above the exact values
for $n=0$ and $1$, and from the negative into the positive for $n\ge 2$.
}
\label{VARfig}
\end{figure}

To calculate $\vev{H}_n^{(1)}$ we need, in addition to \eref{HSterms},
\begin{eqnarray}
\lefteqn{\braket{n^{(1)}}{n^{(1)}}}
\vphantom{\Big|}\cr
& = & \dfrac{(n+1)(n+2)}{4(2n+3)^2}
\bigg(\dfrac{Z(1-X)}{4}-\dfrac{2n+3}{8}\bigg)^2
+\dfrac{(n+1)(n+2)(n+3)(n+4)}{2^{12}(2n+5)^2}
\vphantom{\Bigg|}\cr
& & 
+\dfrac{n(n-1)}{4(2n-1)^2}
\bigg(\dfrac{Z(1-X)}{4}-\dfrac{2n-1}{8}\bigg)^2
+\dfrac{n(n-1)(n-2)(n-3)}{2^{12}(2n-3)^2}
\vphantom{\Bigg|}\;.
\end{eqnarray}
The graphs of $\vev{H}_n^{(0)}$ and $\vev{H}_n^{(1)}$
as well as their stationary points are shown for the $X=0$ case in \fref{VARfig}
for $n=0$ through $n=5$.
Comparing the $Z$-dependence of $\vev{H}_n^{(1)}$ (dashed line) and $\sum_{j=0}^{3}E_n^{(j)}$ (dotted line), 
we can see that they both have three stationary points: two local minima and one local maximum.

For the $n=0$ (ground state, even parity) and $n=1$ (1st exited state, odd parity) cases, 
the global minimum of $\vev{H}_n^{(1)}$ will provide the best approximation without undershooting
the exact values. Indeed, comparing the graphs of 
$\vev{H}_n^{(1)}$ (dashed line) and $\sum_{j=0}^{3}E_n^{(j)}$ (dotted line) for these cases in \fref{VARfig},
we can see that the right local minimum has been lifted to just above the exact value.
These global minima of $\vev{H}_0^{(1)}$ and $\vev{H}_1^{(1)}$ compared to the exact values are
\begin{eqnarray}
\dfrac{\vev{H}_{0,\text{min}}^{(1)}}{E_{0,\text{exact}}} \,=\, 1.00076\;,
\qquad
\dfrac{\vev{H}_{1,\text{min}}^{(1)}}{E_{1,\text{exact}}} \,=\, 1.00066\;,
\vphantom{\Bigg|}
\end{eqnarray}
which are amazingly accurate (better than 0.1\%) for a 3rd order perturbative calculation
without any small expansion parameter.

\begin{figure}[t]
\centerline{
\includegraphics[height=4cm]{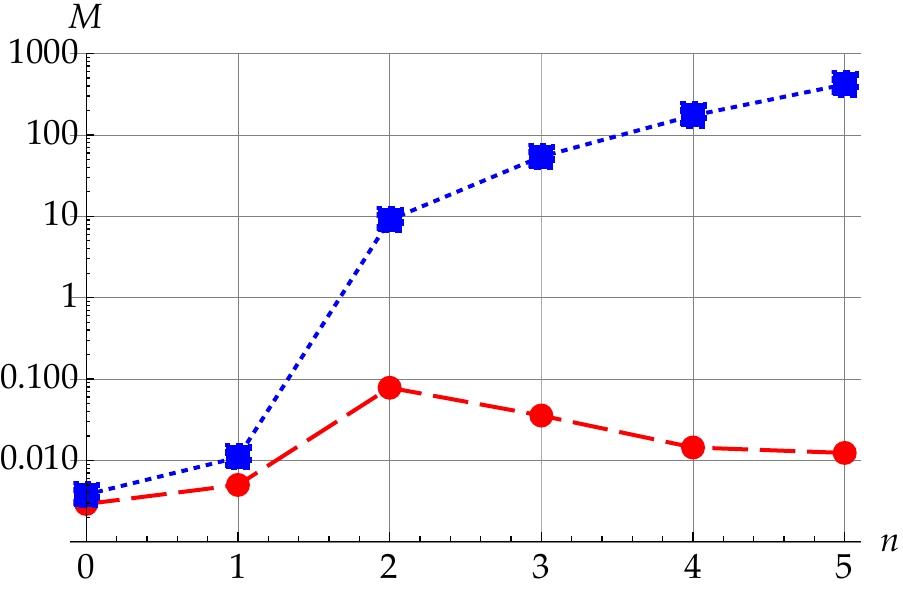}
}
\caption{The spread of $\vev{H}_n^{(1)}$ (circles, dashed) and 
$\sum_{j=0}^{3}E_n^{(j)}$ (squares, dotted) between their respective local minima.
}
\label{EHspread}
\end{figure}

For the $n\ge 2$ cases, the global minimum of $\vev{H}_n^{(1)}$ does not have any special meaning.
Indeed, for the $n=2$ case we can see from \fref{VARfig} that the global minimum of $\vev{H}_2^{(1)}$ is a poor approximation.
Nevertheless, for all $n\ge 2$ cases the right local minimum is now positive, and the spread of $\vev{H}_n^{(1)}$
in the range between the two local minima are greatly reduced compared to that of $\sum_{j=0}^{3}E_n^{(j)}$.
This is shown in \fref{EHspread}.
For $n\ge 3$, the function $\vev{H}_n^{(1)}$ is flat enough between the two local minima, so it does
not matter what value of $Z$ is chosen in that range.

As we continue this procedure to higher orders, we conjecture that $\vev{H}_n^{(k)}$ will become flatter
for a wider range of $Z$ with increasing $k$.
For instance, the next function to consider in the sequence is 
\begin{eqnarray}
\vev{H}_n^{(2)}
& \equiv &
\dfrac{\big(\bra{n^{(0)}}+\bra{n^{(1)}}+\bra{n^{(2)}}\big)\big(\hat{H}_0+\hat{V}\big)\big(\ket{n^{0}}+\ket{n^{(1)}}+\ket{n^{(2)}}\big)}
{\big(\bra{n^{(0)}}+\bra{n^{(1)}}+\bra{n^{(2)}}\big)\big(\ket{n^{0}}+\ket{n^{(1)}}+\ket{n^{(2)}}\big)}
\vphantom{\Bigg|}\cr
& = & E_n^{(0)} + E_n^{(1)}
+ E_n^{(2)}
\dfrac{1 + \braket{n^{(1)}}{n^{(2)}}+\braket{n^{(2)}}{n^{(1)}})}
      {1 + \braket{n^{(1)}}{n^{(2)}}+\braket{n^{(2)}}{n^{(1)}}+\braket{n^{(2)}}{n^{(2)}}}
\vphantom{\Bigg|}\cr
& & +\;
\dfrac{E_n^{(3)} + E_n^{(4)} + E_n^{(5)}}
      {1 + \braket{n^{(1)}}{n^{(2)}}+\braket{n^{(2)}}{n^{(1)}}+\braket{n^{(2)}}{n^{(2)}}}
\;,\vphantom{\Bigg|}
\label{H2def}
\end{eqnarray}
which we expect to have three local minima and two local maxima.
The ``wiggles'' in $\vev{H}_n^{(2)}$ between these extrema should be smaller than those of $\vev{H}_n^{(1)}$.

\section{Discussion}

We have studied the $Z$-value selection problem in
the Halliday-Suranyi approach to the quartic anharmonic oscillator.\cite{Halliday:1979vn,Halliday:1979xh}
We have analyzed the pure quartic potential case ($m=0$) and found that 
the FAC and PMS methods lead to fractional errors that decrease monotonically with increasing $n$.
This is in stark contrast to usual perturbation theory in which the energies of the higher excited
states are more difficult to calculate.

The method can be improved by replacing the $(2k+1)$st order energy $\sum_{j=0}^{2k+1}E_n^{(j)}$
with the expectation value of $\hat{H}=\hat{H}_0+\hat{V}$ for the $k$th order state $\sum_{j=0}^{k}\ket{n^{(j)}}$.
This expectation value is positive definite, and for the $n=0$ (ground state) and $n=1$ (1st excited state) 
cases bounded from below by the exact energies.
At $k=1$ the resulting approximations for the $n=0$ and $n=1$ energies are better than 0.1\%.
The dependence on the value of $Z$ is also greatly reduced compared to $\sum_{j=0}^{2k+1}E_n^{(j)}$,
facilitating the choice of $Z$.

While this result is quite interesting in itself, the more important question is whether
analogous techniques can be applied to $d+1$ dimensional QFT.
Suggestions exist in the literature\cite{Weinstein:2005kx} but the details need to be worked out.

\section*{Acknowledgements}

TT thanks P. Suranyi for his friendship over the years and
L. C. R. Wijewardhana for inviting him to contribute to \textit{Peter Suranyi Festschrift}.
Helpful discussions with D. Minic and C. H. Tze are gratefully acknowledged. 
TT and NB are supported in part by the U.S. Department of Energy (DE-SC0020262, Task C). 

\newpage
\bibliographystyle{ws-procs961x669}
\bibliography{Suranyi}

\end{document}